\documentclass[11pt]{article}
\usepackage{latexsym}
\newcommand{\beq}{\begin{equation}}
\newcommand{\eeq}[1]{\label{#1}\end{equation}}
\newcommand{\bea}{\begin{eqnarray}}
\newcommand{\eea}[1]{\label{#1}\end{eqnarray}}
\begin{document}
\baselineskip 18pt
\begin{titlepage}
\hfill hep-th/0409172
\vspace{20pt}
\begin{center}
{\large\bf MASSIVE GRAVITY IN ADS AND MINKOWSKI BACKGROUNDS}
\end{center}

\vspace{6pt}

\begin{center}
{\large M. Porrati} \vspace{20pt}

{\em Department of Physics\\ New York University\\ 4 Washington Place\\ 
New York NY 10003, USA}

\end{center}

\vspace{12pt}

\begin{center}
{\bf Abstract} 
\end{center}
\begin{quotation}\noindent
I review some interesting features of massive gravity in two maximally 
symmetric backgrounds: Anti de Sitter space and Minkowski space. 
While massive gravity in AdS can be seen 
as a spontaneously broken, UV safe theory, no such interpretation exists yet
in the flat-space case. Here, I point out the problems encountered in trying 
to find such completion, and possible mechanisms to overcome them. 
 
\end{quotation}
\vfill
 \hrule width 5.cm
\vskip 2.mm
{\small \noindent e-mail: massimo.porrati@nyu.edu}
\end{titlepage}
\section{Introduction}
While doing theoretical research in gravity, both 
classical and quantum, standard or 
``super,'' at some stage we inevitably encounter a major contribution due to 
Stanley Deser.
My experience is no exception: in~\cite{bd}, Boulware and Deser 
presented a 
comprehensive and in some way definitive study of massive gravity in four 
dimensions. Their analysis pointed out the true reason underlying the 
problems faced by a quantum theory of massive gravity. 
It has nothing to do with the
incompleteness of Einstein gravity at high energy. Rather, it is a truly 
infrared problem. The problem is best explained using the ADM 
formalism~\cite{adm}, 
another major contribution to gravity due to Stanley.
 
In ADM, the physical, propagating degrees of freedom of massless gravity are
the space metric, $g_{ij}$ $i=1,2,3$ and its conjugate momenta $\pi^{ij}$. 
The other components of the metric, $N=(-g^{00})^{-1/2}$ and $N^i=g^{i0}$, are
nondynamical and appear linearly in the Einstein action. So, they act as 
Lagrange multipliers, enforcing 4 extra constraints. They, together with
the 4 gauge invariances following from general covariance, remove 8 of the
6+6 degrees of freedom, leaving only 2 propagating degrees of freedom 
(2 generalized  
coordinates and 2 conjugate momenta). A Lorentz invariant mass term does not
change the fact that $N^i,N$ are nondynamical, but makes them appear 
nonlinearly in the action of massive gravity. So, their equations of motion
do not produce any new constraint, and one ends up with 6 propagating 
degrees of freedom. One of them is always either a ghost or a tachyon. The 
only exception to this conclusion obtains when Einstein's action 
is modified by adding a 
Pauli-Fierz~\cite{pf} mass term that, at quadratic order in the metric 
fluctuation $h_{\mu\nu}=g_{\mu\nu}-\eta_{mn\nu}$, reads
\beq   
S_{M}= S_{Einstein} + {M^2\over 64\pi G}\int d^4x (h_{\mu\nu}h^{\mu\nu}-h^2).
\eeq{d1}
In this action, the lapse (better $h_{00}\approx N^2-1$) appears linearly, so
that it still acts as a multiplier, and does eliminate the unwanted sixth 
degree of freedom. On the other hand, Boulware and Deser showed that this 
property of the Pauli-Fierz mass term holds only in the quadratic 
approximation. In any Lorentz-invariant mass term, $N$ does enter nonlinearly 
in the complete, interacting Lagrangian. Correspondingly,  
the full nonlinear theory propagates 6 degrees of freedom (and the
Hamiltonian is unbounded below). 

The results of Boulware and Deser mean that massive gravity cannot be a 
consistent quantum theory. The recent revival of interest in massive gravity, 
or, more generally, in long-distance modifications to gravity, does not 
contradict that. The question posed in recent years is not whether a
quantum theory of massive gravity exists, that makes sense up to a very high 
(Planckian) energy. The question is instead whether massive gravity makes 
sense as a {\em low-energy effective field theory}, up to the shortest scale
at which we have experimentally tested gravity. From this point of view, the
UV cutoff of the theory is not $M_{Pl}$ but the (somewhat smaller!) scale
$\sim (100\, \mu m)^{-1}\sim 10^{-3} \, eV$. 
Surprisingly, finding an effective theory that works up to such a small cutoff
and is compatible with experiment is nevertheless 
 difficult. Let us review the 
problems, starting with the case that is better understood theoretically, 
namely, massive gravity in AdS space.
\section{Massive Gravity in AdS Space}
This section is based on~\cite{p1,p2}. 

Neither in flat space, nor in anti de Sitter space is the long-distance 
behavior of Einstein's gravity changed by coupling it to massive particles. 
The effect of massive particles is always encoded in local operators that do 
not give a mass term to the graviton, because of general covariance. 
On the other hand, in AdS space, the effect of massless particles is 
subtler. 
Let us take for instance a very simple case: a conformally coupled free
scalar. Free means here that the scalar interacts only with gravity. By 
integrating out the scalar field, one gets a {\em nonlocal} action for the
graviton, that can be written schematically as
\beq
S={1\over 16\pi G}\int d^4x \sqrt{-g}(R-2\Lambda) + W_{CFT}[g].
\eeq{d2}
$\Lambda$ is the (negative) cosmological constant of the background, and
$W_{CFT}[g]$ denotes the generating functional of the connected correlators
of the CFT. Denote by $\bar{g}_{\mu\nu}$ the background, and expand the 
metric as $g_{\mu\nu}=\bar{g}_{\mu\nu} + h_{\mu\nu}$. Then the 
linearized equations of motion obtained by varying the action in 
Eq.~(\ref{d2}) are
\beq
L_{\mu\nu}^{\alpha\beta}h_{\alpha\beta}+ \Sigma_{\mu\nu}^{\alpha\beta}
h_{\alpha\beta}=0.
\eeq{d3}
Here, $ L_{\mu\nu}^{\alpha\beta}$ is the standard Einstein kinetic operator in
AdS, and 
\beq
\Sigma_{\mu\nu}^{\alpha\beta}=\left.{\delta^2 W_{CFT}\over \delta g^{\mu\nu}
\delta g_{\alpha\beta} } \right|_{g=h}=\langle T_{\mu\nu} T^{\alpha \beta}
\rangle_{CFT}.
\eeq{d4}
So, $\Sigma$ is the two-point function of the stress-energy tensor in the CFT.
Implicit in this notation is the fact that while the CFT is integrated out 
exactly, gravity is treated classically, i.e. all graviton loops are being 
ignored. This approximation makes sense for computing infrared quantities
on a weakly curved background.

By a wise choice of counterterms, $\Sigma$ can be made transverse traceless 
with respect to the background metric. By construction, the equations of 
motion are covariant. So, one can decompose the metric fluctuation into  
transverse-traceless (TT), longitudinal, and trace components, and 
choose the gauge $h_{\mu\nu}=h^{TT}_{\mu\nu}+ \bar{g}_{\mu\nu} \phi$. In this
gauge the equations of motion split into
\bea
\left[{1\over 32\pi G} (\Delta -2\Lambda) + \Sigma(\Delta) \right] 
h_{\mu\nu}^{TT}&=&0 \nonumber \\
(3\Delta -4\Lambda )\phi&=&0.
\eea{d5}
Here, $\Delta$ is the Lichnerowicz operator~\cite{l}, a curved-space 
generalization of the Laplacian, that commutes with the  
covariant divergence and trace defined in terms of the background metric. 
The scalar operator $\Sigma(\Delta)$, 
computed at the pole of the propagator, gives the graviton square mass. In the 
Einstein theory, the pole is at $\Delta=2\Lambda$, and $\Sigma(2\Lambda)=0$.
When the mass is smaller than $2|\Lambda|$, this prescription gives
\beq
M^2 \approx 32\pi G\Sigma(2\Lambda).
\eeq{d6}
Now, even before doing any explicit computation, it is clear that this mass is
parametrically smaller than the curvature radius of AdS, $L\equiv 
\sqrt{|\Lambda|/3}$. Indeed, $\Sigma$ is computed 
by a correlator in the CFT, that depends on $L$ but not on the Newton constant
$G$. So, $\Sigma$ 
can be at most $O(cL^{-4})$, where $c$ is the central charge of the CFT; thus,
\beq
M\sim a\sqrt{c} L_{Planck}/L^2 \ll 1/L.
\eeq{d7}
Here $a$ is a number of order one. Of course, 
it could still be zero. 
The analysis performed in ref.~\cite{p1} shows that $a$ indeed vanishes 
when the conformally coupled scalar is given standard (reflecting) boundary
conditions at the boundary of AdS. When the field is given more general 
boundary conditions, that allow for an energy flow into and out of the AdS 
space, then $a$ is nonzero. 

More precisely, the representation theory of the isometry group of $AdS_4$,
$SO(2,3)$, shows that a conformally coupled scalar can belong to two 
representations, called $D(1,0)$ and $D(2,0)$, respectively 
(see~\cite{p1} or~\cite{h} for notations and further results). When 
reflecting boundary conditions are given, then the scalar belongs to either
$D(1,0)$ or $D(2,0)$~\cite{bf}. In general, the scalar field may be a linear 
combination of modes belonging to both representations. In the case that the
linear combination is the same for all modes, then the scalar propagator is
\beq
\Delta(x,y)= \alpha \Delta^{1}(x,y) + \beta \Delta^{2}(x,y), \qquad 
\alpha+\beta=1.
\eeq{d8}
Here $ \Delta^{E}(x,y)$ is the propagator for modes in the $D(E,0)$ irrep.
Transparent boundary conditions, first proposed in~\cite{ais}, and natural from
the point of view of the holographic AdS/CFT duality~\cite{br,p1}, are 
$\alpha=\beta=1/2$. The result of~\cite{p1,p2} is 
\beq
\Sigma(2\Lambda)=\alpha\beta {G\over 5\pi L^4}.
\eeq{d9}
So, the graviton mass vanishes for standard boundary conditions, but it can 
be nonzero for nonstandard ones. Physically, the Higgs field (a vector 
belonging to $D(4,1)$~\cite{p1}, in this case) is a composite field, a 
``bound state'' of size $L$.

In conclusion, in AdS one can give a (tiny) mass to the graviton by a Higgs 
mechanism. The Higgs is a composite vector of size $L$, and $1/L$ is the cutoff
for the effective theory of massive gravity. Above that energy, the correct
description is in terms of ordinary gravity plus all the degrees of freedom
of the CFT.
\section{Minkowski Space}
No analog of the Higgs-like mechanism described in the previous section is
available in Minkowski space. The best one can do is to introduce the 
appropriate Goldstone fields needed to make massive gravity explicitly
covariant under
general coordinate transformations. This can be done at the full nonlinear 
level~\cite{ahgs}. In the Goldstone boson language, the disease first noticed 
by Boulware and Deser, that the lapse starts propagating at nonlinear level, 
manifests itself in a different guise: massive gravity becomes strongly 
interacting at an extremely small energy scale:
\beq
E\sim (M_{Pl}M^4)^{1/5}, \qquad {\rm or},\qquad
E \sim (M_{Pl}M^2)^{1/3}.
\eeq{d10}
The first scale holds for the standard PF mass term, while the second is the 
best that can be achieved by judiciously improving the action by adding 
appropriately chosen higher-order operators~\cite{ahgs}.
For a graviton of Compton wavelength $M^{-1}\sim 10^{28}\,cm$, the cutoff 
length corresponding to the highest energy cutoff in Eq.~(\ref{d10}) is 
$O(1000\,km)$!
This is the scale below which massive gravity becomes strongly interacting and
essentially uncontrollable within perturbation theory.

The existence of such very low scale is not confined to the theory of a 
single massive graviton. A similar bound also exists in the DGP 
model~\cite{dgp}. The DGP model is the first ghost-free example
of a mechanism in which gravity can be localized on a 4d brane
in a space of infinite transverse volume.
It describes a theory where 4d general
covariance is unbroken, but the graviton is a metastable state. Its main
property is that, on the 4d brane, gravity looks 4d at short distance,
while it weakens at large distance. 
Interesting cosmological applications of this scenario have been proposed, for
instance in~\cite{ddg}.

The model can be described by the action
\bea
S_{\rm DGP} &=& {1\over 16\pi G_5} \int_{{\cal M}} d^5 x\, \sqrt{-g}\, R(g)
\\
&&+ \int_{\partial{\cal M}} d^4 x\, \sqrt{-\gamma} \left[
-{1\over 8\pi G_5} K(\gamma) + {1\over 16\pi G} R(\gamma) \right],
\eea{d11}
where ${\cal M}$ is a 5d manifold with boundary $\partial{\cal M}$,
$g$ is the 5d metric, $\gamma$ is the 4d induced metric on the boundary,
and $K$ is the extrinsic curvature.
The DGP model is closely related to massive gravity.
In fact, the brane-to-brane graviton propagator can be written as
\beq
D_{\mu\nu\,\rho\sigma}^{\rm DGP}(p)
= D_{\mu\nu\,\rho\sigma}^{\rm massive}(p, |p|/L),
\eeq{d12}
where $D^{\rm massive}_{\mu\nu,\lambda\rho}(p, m^2)$ is the propagator for
4d massive gravity, and the ratio of the two Newton constants,
$L=G_5/G$ defines the transition length from standard 4d gravity to the 5d
behavior.

The analysis of~\cite{lpr} (see also~\cite{ddgv}) shows that the DGP model 
becomes perturbatively strongly coupled at a scale $E=(M_{Pl} L^{-2})^{1/3}$.
This is what one would get by naively 
substituting the ``running mass'' $|p|/L$ into the first of the 
bounds in Eq.~(\ref{d10}). Ref.~\cite{lpr} also 
shows that this problem cannot be 
cured by adding local counterterms to the action Eq.~(\ref{d11}).

So, strong coupling at an unacceptably low scale seems an ubiquitous problem 
plaguing IR-modified gravity. It could be resolved, perhaps, 
by a nonperturbative 
re-summation of Feynman diagrams, or by introducing other degrees of freedom
that change the theory before it becomes strongly coupled. In the rest of this
paper, I will briefly discuss the second possibility: the only one that would 
give us full computational control of the theory. 

For simplicity, I will 
discuss massive gravity~\footnote{Refs.~\cite{pr,gv} are a first attempt to 
studying these issues in the DGP model.}.
The strong coupling problem stems from the breakdown of the linearized 
approximation. This happens at a lower than expected scale because the 
linearized graviton fluctuation generated by a conserved source with stress 
energy tensor $T_{\mu\nu}$ is (in momentum space):  
\bea
h_{\mu\nu}(p)&=& \bar{h}_{\mu\nu}(p) + p_\mu p_\nu \Psi(p), \nonumber \\
\bar{h}_{\mu\nu}&=&
{8\pi G  \over p^2 + M^2 }\left[ T_{\mu\nu}(p) -{1\over 3}T_{\mu\nu}(p)\right],
\qquad \Psi(p)=-{1\over 3M^2} {8\pi G  \over p^2 + M^2 } T_\mu^\mu (p).
\eea{d13}
The first term, $\bar{h}_{\mu\nu}$ is well-behaved in the limit 
$M\rightarrow 0$; the factor $-1/3$ in its trace component is the origin of the
famous van Dam-Veltman-Zakharov discontinuity~\cite{vdvz}, and it also
ensures that 
the only propagating degrees of freedom are the 5 physical polarizations of a 
massive spin-2 field.
On the other hand, $\Psi$ diverges in the massless limit. 
At liner order, this infrared divergence is harmless, since $\Psi$ is a  
gauge mode: it vanishes in the one-graviton scattering amplitude, when 
$h_{\mu\nu}(p)$ is contracted with a conserved source. At the next order, 
though, it contributes an amplitude that may dominate over the linear term. For
a point-like source of mass ${\cal M}$, 
inspection of Eq.~(\ref{d13}) shows that 
this happens at distances $r=O(G{\cal M}/M^4)$. The length scale 
$(M^4 M_{Pl})^{-1/5}$ is attained for ${\cal M}=M_{Pl}$.

Can we modify massive gravity in such a way as to keep the same one-graviton
amplitude in between conserved sources as given by Eq.~(\ref{d13})? The 
answer is: yes, by modifying the graviton's {\em kinetic} term.

Let us add to the linearized action $S_{M}$ in Eq.~(\ref{d1}) the term
\beq
S_A= {A\over 32\pi G}
\int d^4x h (\partial_\mu \partial_\nu h^{\mu\nu} - \Box h),
\eeq{d14}
where $A$ is an arbitrary constant. 

Decompose next the metric into its transverse-traceless, longitudinal, and 
trace components: 
\beq
h_{\mu\nu}= h_{\mu\nu}^{TT} + \eta_{\mu\nu} \Phi + 
\partial_\mu \partial_\nu \Psi + \partial_{(\mu}A^T_{\nu)}, \qquad 
\partial^\mu A^T_\mu =0.
\eeq{d15}
By computing the double divergence of the equations of motion  
($\delta (S_{M} + S_A)/\delta h_{\mu\nu}-T_{\mu\nu}=0$) we get
\beq
(M^2 \Box + A\Box^2) \Phi =\partial_\mu\partial_\nu T^{\mu\nu}.
\eeq{d16}
So, even when  $A\neq 0$, $\Phi$ still obeys a homogeneous equation when 
$T_{\mu\nu}$ is conserved, and so it can be set to zero when computing 
the field generated by a localized source.
This is the key property that guarantees that only 5 physical polarizations 
propagate in the one-graviton scattering amplitude.
The unphysical gauge mode $\Psi$ changes. It becomes
\beq
 \Psi(p)=-{1\over 3(M^2 + Ap^2)} {8\pi G  \over p^2 + M^2 } T_\mu^\mu (p).
\eeq{d17}
For all $p^2\neq 0$, this mode can be made arbitrarily small in the limit
$A\rightarrow \infty$~\footnote{Static sources and quantum loop computations 
both require to use Euclidean momenta, so this condition is generic.}. 
Formally, the limit $A\rightarrow \infty$ eliminates the dangerous mode that
triggers the breakdown of the linear approximation. Moreover, when expanding 
the metric as in Eq.~(\ref{d15}), the change in the kinetic term can be easily
interpreted as giving a very large kinetic term to the unwanted modes
$\Phi$ and $\Psi$, that, therefore, decouple.

This argument is of course still hand-waving and it would be interesting to 
make it sounder. One objection that can be raised against it is that it makes
$h_{\mu\nu}$ propagate 7 degrees of freedom, two of which are ghosts, 
instead of the physical 5. 
So, for any finite $A$, the theory is unstable unless 
the ghosts decouple from all physical amplitudes, as the do at linear order.

The existence of two extra degrees of freedom is another simple 
application of the
methods devised by Boulware and Deser~\cite{bd}. By writing the action in terms
of the 3d metric, linearized lapse $h_{00}$, and shift $h_{0i}$, we see that
$S_A$ is
\beq
S_A={A\over 16\pi G} \int d^4x (h_{ii} -h_{00})\partial_i 
\partial_0h_{i0} +...
\eeq{d18}
So, $\partial_i h_{i0}$ and $h_{ii} -h_{00}$ become a new
pair of (propagating) canonical variables. Now the total number of degrees of
freedom is thus 6 (coming from $h_{ij}$) plus 1. The new degree of freedom
is a boson with first-order action in the time derivative, so its energy is
unbounded below.

Whether massive gravity --or DGP, where a more sophisticated version of this 
mechanism may be at work, according to the analysis of~\cite{pr}-- can be
made perturbatively stable and calculable is yet to be proved, many years 
after the groundbreaking investigations of Boulware and Deser. The problem 
is still tantalizing, so much so that it may be fit to conclude this review by
mentioning that another intriguing route to solve the strong coupling 
problems of gravity has been recently opened: by explicitly breaking Lorentz
invariance, gravity can be made finite-range, consistent with existing data,
and weakly coupled down to distances $O(100\,\mu m)$.~\cite{ahclm,r,d}
\subsection*{Acknowledgments}
Work supported in part by NSF grant PHY-0245068. 

\end{document}